# AQUILA: A Laboratory Facility for the Irradiation of Astrochemical Ice Analogues by keV Ions

R. Rácz,[1,†] S.T.S. Kovács,[1] G. Lakatos,[1,2] K.K. Rahul,[1] D.V. Mifsud,[1] P. Herczku,[1] B. Sulik,[1] Z. Juhász,[1] Z. Perduk,[1] S. Ioppolo,[3] N.J. Mason,[1,4] T.A. Field,[5] S. Biri,[1,†] and R.W. McCullough[5,†]

1. HUN-REN Institute for Nuclear Research (Atomki), Debrecen H-4026, Hungary
2. Institute of Chemistry, University of Debrecen, Debrecen H-4032, Hungary
3. Centre for Interstellar Catalysis (InterCat), Department of Physics and Astronomy, Aarhus University, Aarhus DK-8000, Denmark
4. Centre for Astrophysics and Planetary Science (CAPS), School of Physics and Astronomy, University of Kent, Canterbury CT2 7NH, United Kingdom
5. Department of Physics and Astronomy, School of Mathematics and Physics, Queen's University Belfast, Belfast BT7 1NN, United Kingdom

†   Corresponding authors:   R. Rácz (rracz@atomki.hu); S. Biri (biri@atomki.hu); R.W. McCullough (rw.mccullough@qub.ac.uk)

**Abstract**

The detection of various molecular species, including complex organic molecules relevant to biochemical and geochemical processes, in astronomical settings such as the interstellar medium or the outer Solar System has led to the increased need for a better understanding of the chemistry occurring in these cold regions of space. In this context, the chemistry of ices prepared and processed at cryogenic temperatures has proven to be of particular interest due to the fact that many interstellar molecules are believed to originate within the icy mantles adsorbed on nano- and micro-scale dust particles. The chemistry leading to the formation of such molecules may be initiated by ionising radiation in the form of galactic cosmic rays or stellar winds, and thus there has been an increased interest in commissioning experimental set-ups capable of simulating and better characterising this solid-phase radiation astrochemistry. In this article, we describe a new facility called AQUILA (Atomki-Queen's University Ice Laboratory for Astrochemistry) which has been purposefully designed to study the chemical evolution of ices analogous to those that may be found in the dense interstellar medium or the outer Solar System as a result of their exposure to keV ion beams. The results of some ion irradiation studies of $CH_3OH$ ice at 20 K are discussed to exemplify the experimental capabilities of the AQUILA as well as to highlight its complementary nature to another laboratory astrochemistry set-up at our institute.

## 1. Introduction

The discovery of the first molecules in interstellar space about a century ago **[1-5]** ushered in a new paradigm in astrophysics; one that has increasingly recognised the chemical richness of the cosmos. To date, over 300 individual molecules have been detected in interstellar space, along with an additional 70 in extra-galactic sources **[6]**. These molecules include several of relevance to biochemistry, including urea and formamide **[7,8]**, along with others of mineralogical and geochemical importance such as silicon monoxide **[9]**. The relevance of interstellar molecules to the emergence of life and the development of planetary systems has thus provided a strong motivation for better comprehending the astrochemical mechanisms leading to their formation and destruction.

The mechanisms by which molecular formation (and destruction) in interstellar space occurs were summarised by Herbst **[10]** and Arumainayagam *et al.* **[11]**, who described three milieux for astrochemical reactions: (i) chemistry between gas-phase molecules, (ii) reactions catalysed by the surfaces of interstellar dust grains, and (iii) reactions occurring within icy grain mantles. The reactions taking place on the surfaces of dust grains or within icy mantles have long been recognised to be a dominant source of interstellar molecules, particularly within dense interstellar clouds **[12-14]**. The interaction of ionising radiation (largely in the form of galactic cosmic rays, stellar winds, or vacuum-ultraviolet photons) with interstellar ices is of particular interest, as laboratory studies have determined that the energy deposited into the ice as a result of this interaction is able to trigger a cascade of chemical reactions that could result in the formation of a number of complex organic molecules, including those relevant to biology **[15,16]**.

Accordingly, a number of experimental research groups have designed set-ups to allow them to probe this radiation astrochemistry. Historically, the first laboratory astrochemistry apparatus was that based at Leiden Observatory in the Netherlands, which made use of a high-vacuum chamber within which astrochemical ice analogues could be prepared on a substrate cooled to 10 K **[17]**. The ices could then be processed by Lyman-α photons supplied by ultraviolet lamps attached to the chamber. Since this first set-up was established, a number of other laboratories have established their own astrochemical chambers, each offering a different technique of preparing, processing, or analysing astrochemical ice analogues. Among these set-ups are the Keck Machine in Hawaii **[18]**, the VIZSLA set-up in Budapest

[19], the LISA set-up in Nijmegen [20], the CASIMIR and IGLIAS chambers in Caen [21,22], the SURFRESIDE$^2$ set-up in Leiden [23], and the PAC in Milton Keynes [24]; to name but a few.

In this article, we describe a new experimental astrochemistry chamber that has been designed and installed as a permanent end-station of the electron cyclotron resonance (ECR) ion source at the HUN-REN Institute for Nuclear Research (Atomki) in Debrecen, Hungary. The commissioning of this chamber is the result of a strong collaborative relationship between Atomki and Queen's University Belfast in the United Kingdom. This relationship is reflected in the chosen name of the set-up: AQUILA (Atomki-Queen's University Ice Laboratory for Astrochemistry). Indeed, the core components of the AQUILA were originally designed and constructed by R.W. McCullough and A.C. Hunniford at the ECR ion source laboratory at Queen's University Belfast so as to carry out sputtering studies of astrochemical ice analogues induced by the impact of 4 keV singly and multiply charged ions [25,26]. Now established at Atomki, the AQUILA allows us to probe the chemistry induced in astrochemical ice analogues under conditions relevant to the interstellar medium and the outer Solar System by a variety of ions having typical energies of a few tens to a few hundreds of keV.

The commissioning of the AQUILA augments and complements the astrochemical capabilities of the Atomki laboratories, which had already been served by the Ice Chamber for Astrophysics-Astrochemistry (ICA) [27,28]. The ICA is currently installed as a permanent end-station of the Atomki Tandetron accelerator, and may thus be used to study the radiation chemistry of astrochemical ice analogues induced by higher energy ions having energies of a few hundred keV to a few MeV. The availability of lower energy ions supplied by the ECR ion source therefore extends the energy range over which ion-induced radiation astrochemistry studies may be performed at Atomki, allowing for various space radiation sources to be considered such as high-energy galactic cosmic rays and lower energy stellar winds or giant planetary magnetospheric plasmas. Section 2 of this article is devoted to an overview of the Atomki ECR ion source and the manipulation of the ion beams it is able to supply, while Section 3 provides a detailed description of the AQUILA chamber. Results obtained from the H$^+$ and oxygen ion irradiation of CH$_3$OH ice at 20 K are discussed in Section 4, while results obtained from temperature-programmed desorption (TPD) experiments of pristine and irradiated CH$_3$OH ices are provided in Section 5. A detailed comparison of the AQUILA and ICA set-ups is provided in Section 6, and final concluding remarks are given in Section 7.

## 2. Ion Beam Generation and Manipulation

### 2.1 The Atomki ECR Ion Source

Although a detailed description of the Atomki ECR ion source may be found in previously published articles [29-31], a brief overview of its most salient features is provided herein so as to better contextualise our description of the AQUILA set-up. The ECR ion source, which was built in-house and has been in operation since 1996, is able to supply various versatile low-energy ion beams for use in different fields of study, including atomic physics, plasma science, materials science, and, of course, astrochemistry. Ion beams are produced by first generating the plasma state of the required material from its precursor gas *via* GHz frequency microwave injection into the plasma chamber of the ion source. This plasma is confined by a magnetic field that is generated as a result of the superposition of permanent magnets and electrical solenoids. The efficient energy transfer from the microwave to the plasma electrons allows for singly or multiply charged ions to be readily produced, with high-intensity beams (i.e., having currents of at least a few tens or even a few hundred of μA) of H$^+$, He$^{q+}$, C$^{q+}$, N$^{q+}$, O$^{q+}$, Ne$^{q+}$, Si$^{q+}$, S$^{q+}$, Ar$^{q+}$, Kr$^{q+}$, and Xe$^{q+}$ ions being routinely produced [31].

Ion beams may also be produced from solid materials by making use of various vaporisation techniques, such as sputtering, sublimation, or oven-heating. In this way, singly or multiply charged P$^{q+}$, Ca$^{q+}$, Fe$^{q+}$, Ni$^{q+}$, Ag$^{q+}$, Au$^{q+}$, and C$_{60}^{q+}$ ion beams have been successfully prepared. Indeed, the highest charge-state ion beam that has been produced was a 900 keV $Au^{30+}$ beam. Aside from these positively charged atomic beams, the ion source is also able to provide stable beams of positively charged molecular ions (e.g., $H_2^+$, $H_3^+$, $OH^+$, $H_2O^+$, $H_3O^+$, and $O_2^+$), as well as various negatively charged ion beams (e.g., $H^-$, $O^-$, $OH^-$, $O_2^-$, $C^-$, $C_{60}^-$) having beam currents of at least a few tens of µA and a few µA, respectively [30].

The kinetic energy of the ion beams supplied by the ECR ion source, $E_{beam}$, is related to the charge-state of the projectile ion, $q$, as follows:

$$E_{beam} = qV_{ext}$$

(1)

where $V_{ext}$ is the extraction voltage, which may be set to a value between 50 V and 30 kV. Given that it is possible to produce ions with charge-states ranging between $q$ = 1-30, this means that the ECR ion source is able to supply ion beams that cover the 50 eV to 900 keV energy range. However, for experiments on the radiation astrochemistry of ices, extraction voltages of 5-20 kV are typically used, meaning that 5-20 keV singly charged ions, as well as multiply charged ions having energies of up to a few hundred keV are routinely available for laboratory astrochemistry experiments.

## 2.2    *The Ion Beam Transport Line*

The ECR ion source is connected to two beamlines (Figure 1). The original 90° beamline is used for testing purposes, for studies in atomic physics, and for the functionalisation of medical or industrial samples *via* ion implantation. A recently developed 55° beamline connects the ion source to the AQUILA set-up, and is thus dedicated to ion irradiation studies of astrochemical ice analogues. The pathway that an ion beam takes from the ECR ion source to the AQUILA set-up *via* this 55° beamline is summarised in Figure 2. Once the ion beam leaves the ion source, it may be focused or de-focused using two einzel lenses before being passed through the 55° bending magnet. The emergent ion beam is subsequently passed through a collimator having a diameter of 24 mm (labelled C1 in Figure 2), thereby providing the first definition of the beam-spot diameter. A magnetic lens installed just beyond this collimator may then be used to manipulate the beam in the *x*-axis (the direction of travel of the beam being defined as the *z*-axis), after which another einzel lens is used to focus the ion beam through a second collimator (labelled C2 in Figure 2) having a diameter of 15 mm, thus reducing the beam-spot diameter.

The ion beam is subsequently passed through two pairs of deflector plates (labelled DP in Figure 2) which are able to deliver an alternating electric field across the *x*- and *y*-axes supplied by high-voltage ramp generators with respective minimum frequencies of 25 and 600 Hz so as to stabilise and homogenise the current density of the beam. The beam is subsequently passed through a third collimator (labelled C3 in Figure 2) having a diameter of 10 mm which largely determines the beam-spot size supplied to the AQUILA.

Located immediately beyond this collimator is a system composed of a Faraday cup (labelled F1 in Figure 2) and a mesh (labelled M1 in Figure 2), either of which can be inserted into the ion beam pathway at the expense of the other. The stainless-steel Faraday cup, which has an outer diameter of 20 mm and a length of 30 mm, allows for the ion beam current to be quantitatively determined. At the bottom of the cup is a 60° cone which reduces the effect of any secondary electrons that may have been emitted as a result of the interaction of the incident ions with the Faraday cup. During standard current measurement procedures, the Faraday cup is also biased by a voltage of +45 V. Once the current has been adequately measured, the mesh component of the system is inserted into the beam pathway. This 20 mm-diameter mesh is composed of BeCu wires having a thickness of 20 µm and a relative spacing from each other of 0.5 mm, and allows a beam transmission of 90%.

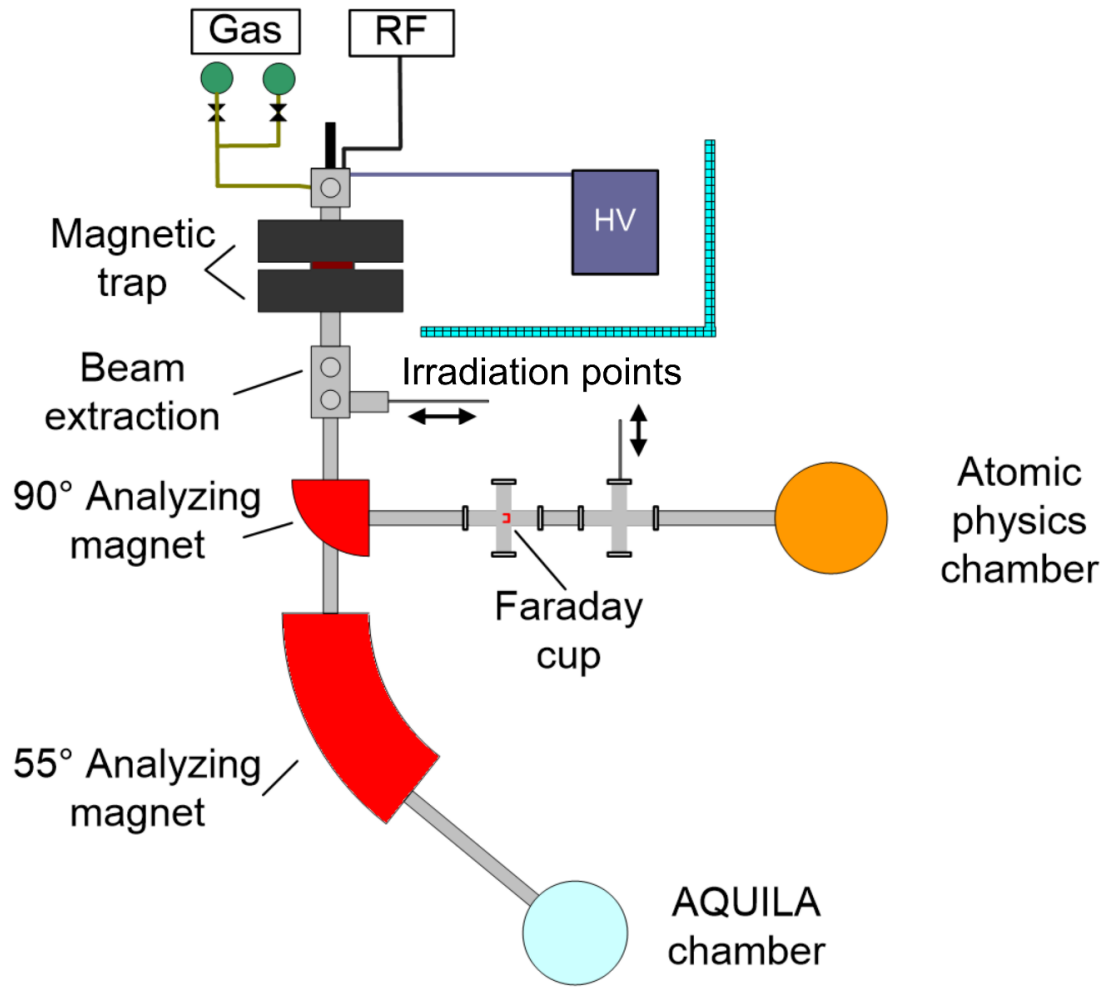

**Figure 1.** A top-view schematic diagram of the Atomki ECR ion source beamlines, which lead to an atomic physics chamber and the AQUILA chamber.

The ultrahigh-vacuum (UHV) AQUILA chamber is separated from the beamline vacuum systems by means of a pneumatic gate valve. Between the gate valve and the entrance port of the chamber, there is a fourth collimator (labelled C4 in Figure 2) having a diameter of 10 mm which serves to trim any scattered ions out of the beam profile. Once the beam has entered the AQUILA chamber, a second system composed of a stainless-steel Faraday cup and a BeCu mesh (respectively labelled F2 and M2 in Figure 2) is used to measure the ion beam current and profile within the chamber. The properties (i.e., materials and dimensions) of this second Faraday cup-mesh system F2, M2 are identical to those of the first system F1, M1.

The nominal area of the beam supplied to the target within the AQUILA chamber approximates a homogeneous circle with a diameter of 12 mm. The homogeneity and beam-spot size obtained at the target using defined sets of beam guiding conditions was confirmed through a series of previous trials making use of mm-marked ZnS fluorescent screens. This allowed the optimum beam guiding conditions for the production of a circularly homogeneous beam-spot of diameter 12 mm to be determined. The ion beam current that is actually incident on the target can be routinely measured throughout the course of an experiment by rotating the target (which is mounted on a sample holder) by 180° and using a 10 mm-diameter hole behind the target as a Faraday cage. Since the current measured at the entrance to the AQUILA chamber by the Faraday cup-mesh system F2, M2 is proportional to the current actually incident with the target, then by knowing the proportionality constant it is possible to continuously monitor the uniform ion beam current density at the target.

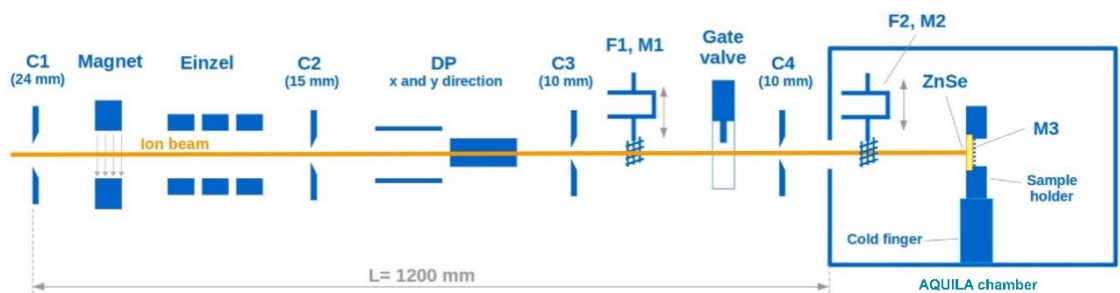

**Figure 2.** Schematic diagram of the controlled ion beam transport line between the 55° bending magnet and the AQUILA chamber (not to scale). More detailed information may be found in the text.

## 3. The AQUILA Set-Up

### 3.1  *Chamber Design and Function*

The AQUILA experimental set-up comprises a cylindrical UHV-compatible stainless-steel chamber (custom-made by Kurt J. Lesker Company) having an inner diameter of 300 mm and a height of 210 mm. The side wall of the chamber is fitted with four DN-100 CF and four DN-40 CF flanges to allow for external connections to, for example, the supplied ion beam or analytical devices (Figure 3). The top and bottom of the chamber are each fitted with a DN-295 CF flange. In the case of the top of the chamber, the centre of this flange is welded to a DN-160 CF reducer as well as an additional two DN-16 CF flanges; while in the case of the bottom of the chamber a DN-100 CF reducer together with four DN-40 CF flanges are welded on.

Within the centre of the chamber is a heat-shielded oxygen-free high-conductivity copper sample holder into which a single infrared-transparent deposition substrate used to prepare astrochemical ice analogues may be mounted (Figure 4). This deposition substrate is typically a ZnSe disc of diameter 15 mm and thickness 3 mm, although other infrared-transparent materials may also be used. Moreover, the deposition substrate may be coated by a thin (< 250 nm) gold-wire meshwork which prevents the accumulation of electrostatic charge and spark release during ion beam irradiation. The gold-wire mesh has a lateral thickness of just 20 μm and a relative separation of 0.8 mm, and thus does not attenuate incident infrared spectroscopic beams to any significant degree **[27]**. Moreover, extensive previous work performed using the ICA has demonstrated that the use of deposition substrates coated with gold-wire meshwork does not influence the resultant structure of a condensed astrochemical ice analogue to any significant degree **[27,28]**.

The sample holder, which may be rotated about the vertical axis of the chamber by means of a 360° rotation stage (Thermionic Northwest Inc., RNN-400), is held in contact with the cold finger of a closed-cycle helium cryostat (Sumitomo CH-204SB-N cold head with a Sumimoto HC-4E1 compressor unit) which is connected to the main AQUILA chamber *via* its bottom flange. This allows the sample holder and the mounted ZnSe deposition substrate to be cooled to a minimum temperature of 20 K, although an operational temperature range of 20-300 K is available and the temperature of the sample holder and substrate may be set to any value within this range by setting an equilibrium between the cooling effect of the cryostat and the warming induced by an internal 40 Ω / 50 W cartridge heater (HTR-25-100). Temperature measurements are performed using three silicon diodes (Lake Shore DT-670B-CO) and a proportional-integral-differential (PID) controller (Lake Shore model 336), and the temperature of the deposition substrate was calibrated by performing TPD experiments using common astrophysical ice analogues and comparing the observed sublimation temperature with literature values. Two of the silicon diodes are connected directly to the sample holder, while the third is connected to the cold finger (Figure 4). At 20 K, the maximum temperature difference recorded by the two diodes connected to the sample holder is typically less than 0.2 K, and so the actual temperature of the sample holder is taken to be a simple average of these two recorded values. Once equilibrated, the temperature of the sample holder may be varied at rates of 0.1-10 K min$^{-1}$.

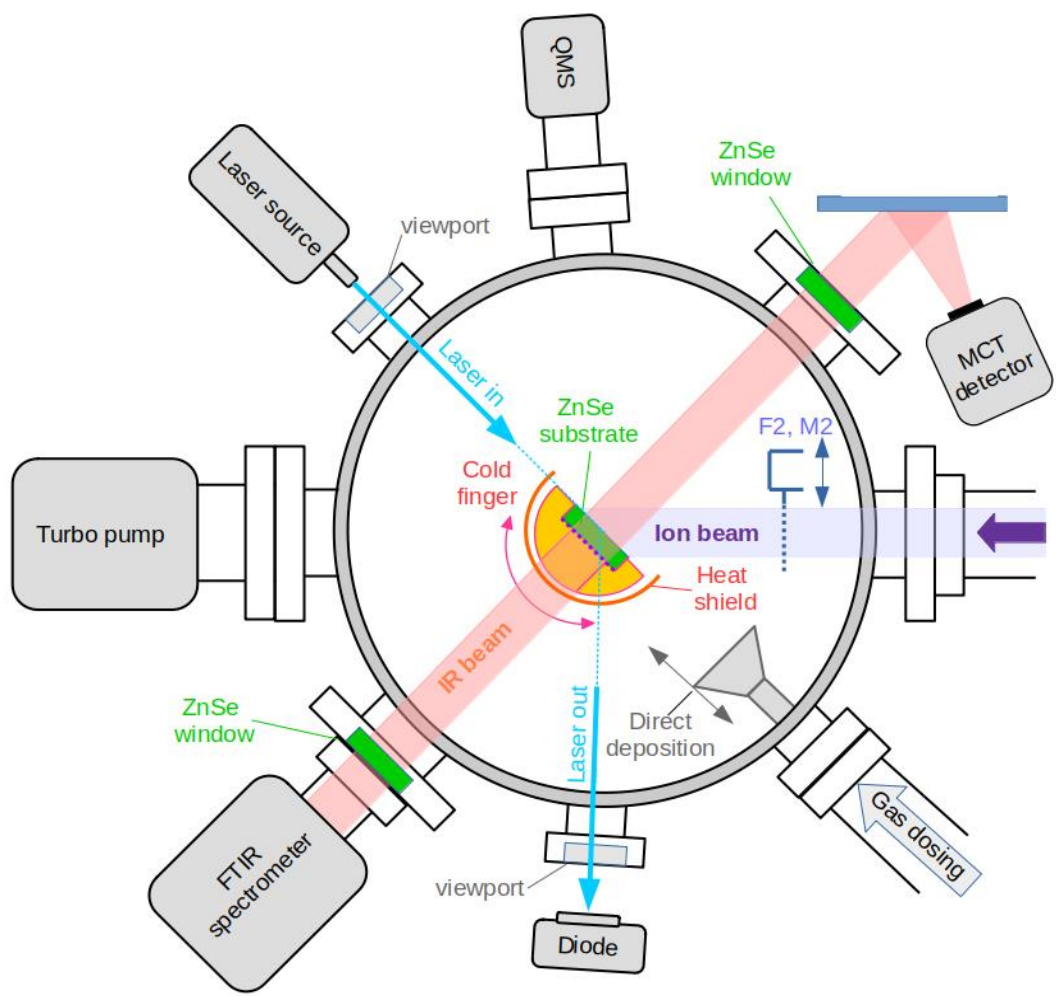

**Figure 3.** A top-view schematic diagram of the AQUILA chamber arranged for transmission absorption mid-infrared spectroscopy. The sample holder and heat shielding are rotatable, which allows for ion beam irradiation experiments to be performed as depicted (i.e., with the infrared beam being held orthogonal to the deposition substrate surface and projectile ions impacting at angles of 45°), as well as for laser interferometry to be used for ice thickness determination when the sample holder is rotated to allow the incident laser beam to be reflected into the photo-diode.

As previously mentioned, the sample holder is surrounded by a heat shield so as to preclude radiative thermal transfer from the walls of the chamber which can raise the temperature of the deposition substrate. As can be seen in Figure 5, the 57-mm diameter heat shield has a 120° opening of height 15 mm in front of the sample holder which allows for the beam supplied by the ECR ion source to interact with the astrochemical ice analogue deposited on the cooled ZnSe deposition substrate. A 15 mm circular hole centred on the ZnSe substrate is also present on the rear side of the heat shield, so as to allow for the transmittance of an infrared spectroscopic beam. Furthermore, a suppressor plate bearing a 10 mm hole has been mounted on the rear side of the heat shield (Figure 5) so as to allow for confirmatory ion beam current density measurements in addition to those described in Section 2: in order to directly measure the current incident on the ZnSe deposition substrate, a mesh system (labelled M3 in Figure 2) may be installed directly behind the substrate to prevent it from charging up. The substrate is then rotated by 180° such that the mesh M3 directly faces the incident ion beam. The current density incident on mesh M3 can be readily measured due to the fact that the sample holder is electrically insulated from the cold head of the cryostat by means of a sapphire rod of length 20 mm and diameter 10 mm.

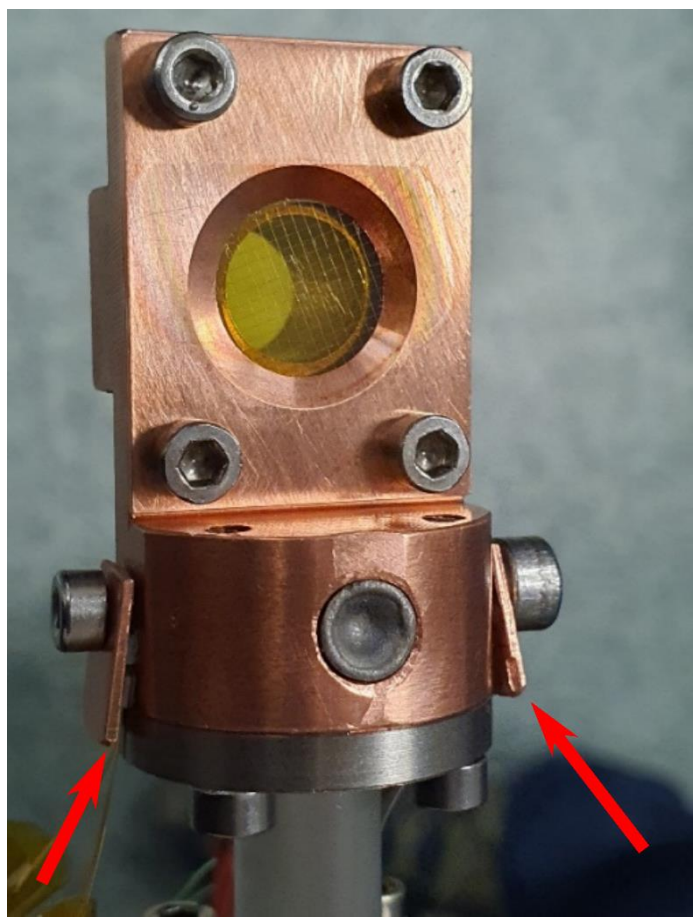

**Figure 4.** Photograph of the oxygen-free high-conductivity copper sample holder within which a single ZnSe deposition substrate has been mounted. It is possible to note the presence of a thin (< 250 nm) gold-wire meshwork on the surface of the ZnSe substrate which serves to preclude the accumulation of electrostatic charge and spark release during ion beam irradiation. Also visible are two of the silicon diodes used to measure the temperature of the sample holder and deposition substrate, indicated by red arrows.

It is important to note that, during ion irradiation experiments of astrochemical ice analogues prepared on ZnSe substrates, no significant increase in the temperature of the deposition substrate (and, by extension, the astrochemical ice analogue deposited on it) is anticipated. This is justified by considering the thermal conductivity of ZnSe which, at 20 K, is known to be in excess of 6 W K$^{-1}$ cm$^{-1}$ **[32]**. By multiplying this value by the surface area of the ZnSe substrate that is exposed to good thermal contact with the copper sample holder and dividing by the estimated maximum possible length of heat transfer through the substrate, it is possible to arrive at a heat conductance of 4.8 W K$^{-1}$ for ZnSe at 20 K. It therefore follows that an incident ion beam with a power dissipation of 1 W will only increase the temperature of the substrate by 0.21 K. As the ion beams used for radiation astrochemistry experiments typically have power dissipations significantly less than 1 W, the resultant heating of the deposition substrate by the ion beam is assumed to be negligible.

Evacuation of the AQUILA chamber (which occurs prior to any cooling of the sample holder and substrate) is achieved using three turbomolecular pumps (two Balzers TMU 260 pumps and a Varian TV301 pump) and one scroll pump (Edwards nX$^{ds}$ 15i) and allows for base pressures of $5\times10^{-9}$ mbar and $10^{-9}$ mbar to be achieved before and after cooling of the sample holder and substrate, respectively. The pressure in the main chamber is measured using a wide vacuum gauge (Edwards WRGS-NW35) and an ionisation gauge (Vacgen ZVIG18) connected to two of the DN-40 CF flanges at the bottom of the chamber. As mentioned previously, the main chamber is separated from the projectile ion beamline (which operates at a base pressure of a few $10^{-7}$ mbar) by a pneumatic gate valve. To better preclude against pollution from the beamline entering the AQUILA, a differential pumping stage composed of two turbomolecular pumps (one Pfeiffer TMH260 and one Edwards EXT250) and a scroll pump (Edwards nX$^{ds}$ 15i) has also been installed. In the event of unsatisfactory base pressure levels within the main chamber, the AQUILA may be baked out using two internal 500 W halogen bulbs (Kurt J. Lesker Company).

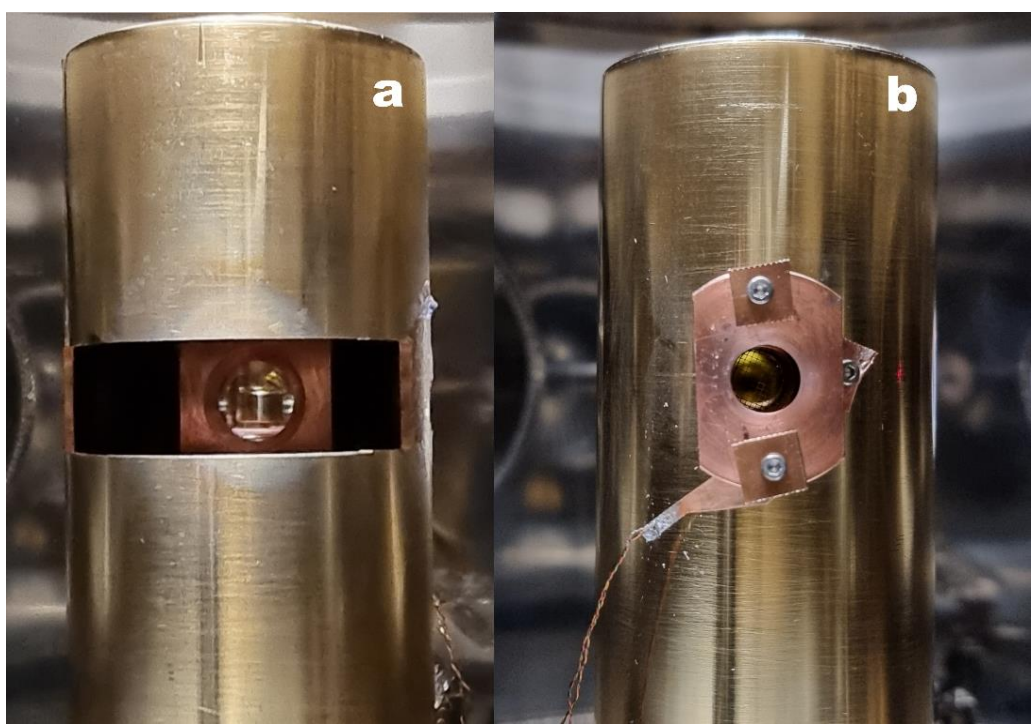

**Figure 5.** Photographs of the heat shield applied to the copper sample holder as viewed from (a) the front and (b) the rear sides. In the case of the latter, it is possible to note the attachment of a suppressor plate to the rear of the heat shield (more detailed information may be found in the text).

### 3.2  *Astrochemical Ice Analogue Preparation, Processing, and Analysis*

The standard experimental protocol involves first evacuating the chamber to base pressure before cooling the sample holder and substrate to the desired cryogenic temperature which, for the remainder of this article, will be assumed to be 20 K. Once thermal equilibrium at this temperature has been reached, an astrochemical ice analogue may be prepared on the infrared-transparent deposition substrate. Ice analogues may be prepared *via* both the background and direct deposition of dosed gases or vapours by making use of two separate pre-dosing mixing containers nominally held under a vacuum of about $10^{-3}$ mbar. Gases from lecture bottles or vapours from vials containing de-gassed liquids may be introduced into these mixing containers using a sophisticated system of valves. In total, four different lecture bottles or vials may be connected to each mixing container at any one time, thus allowing for the preparation of both pure and multi-component astrochemical ice analogues having a defined stoichiometric composition. Contamination of these pre-dosing mixing containers may be precluded by their sequential warming and evacuation prior to the introduction of a gas mixture of interest.

Once introduced into the mixing containers, gas mixtures are allowed to settle for a few minutes to ensure adequate mixing. The total pressure of the gas mixture prior to its introduction into the main chamber is typically on the order of a few mbar and can be measured using mass-independent capacitive manometer gauges (Pfeiffer CCR361 and Pfeiffer CMR361). The actual dosing of the gas mixture into the main chamber is mediated by fine regulating all-metal needle valves (Vacgen M07) and can be performed as either background or direct ice deposition. To facilitate this, two tubes have been welded to the gas inlet port of the chamber (Figure 3); the longer (315 mm) of these tubes is used for direct deposition and, by making use of a 100 mm bellows system, its nozzle may be carefully brought into very close proximity with the surface of the deposition substrate. The shorter (50 mm) tube is used for the background deposition of astrochemical ice analogues, and it possesses a distributor (i.e., scattering) plate mounted directly in front of its nozzle. This distributor plate helps to reduce the pressure inhomogeneity that may arise in the main chamber during gas dosing and therefore allows for an even growth of the depositing astrochemical ice analogue.

The growth of the depositing astrochemical ice analogue can be quantitatively monitored *in situ* using a laser interferometric technique, and thus the AQUILA has been accordingly fitted with a 45 mW laser that emits light having a wavelength of 405 nm, as well as a silicon photo-diode able to detect light across the 305-550 nm wavelength range (Figure 3). This technique, which has been described in detail in other publications [**33,34**], allows for the determination of the thickness of the growing ice layer by measuring the intensity variations of laser light that is reflected off the surface of the deposition substrate during ice deposition. During deposition, a sinusoidal variation in intensity is detected due to the interference of the laser light reflected at the ice-vacuum and ice-substrate boundaries. The number of consecutive pattern repetitions $k$ is related to the thickness of the ice $d$ (μm) as:

$$d = \frac{k\lambda}{2n \cos(\theta)}$$

(2)

where $\lambda$ is the wavelength of the laser light *in vacuo* (μm), $n$ is the refractive index of the ice, and $\theta$ is the reflection angle of the laser light in the ice. This method of determining ice thickness is advantageous as it provides an accurate evaluation of the rate of deposition and only relies upon knowledge of $\theta$ (known from the geometry of the set-up), $\lambda$ (set at 405 nm), and $n$. In the case of this latter term, a large number of previous studies has quantified the refractive indices of various molecular ices [**35-38**], and thus there is a rich literature in which the required data may be found.

The deposition and growth of the ice (as well as its subsequent processing) may also be followed *in situ* using Fourier-transform mid-infrared transmission absorption spectroscopy. The AQUILA is equipped with a Bruker Vertex V70v spectrophotometer and an external mercury-cadmium-telluride detector cooled by liquid nitrogen (Figure 3), which allows studies to be performed over the 4000-650 cm$^{-1}$ (2.5-15.4 μm) range with the mid-infrared spectroscopic beam maintained orthogonal to the plane of the deposition substrate. Spectra may be acquired with a maximum resolution of 0.1 cm$^{-1}$, although lower resolutions of 1 or 2 cm$^{-1}$ are more typically used. The amount of a particular molecular species condensed as an ice onto the deposition substrate may be quantified through its molecular column density $N$ (molecules cm$^{-2}$). This quantity may be calculated from mid-infrared absorption spectra acquired after the deposition of the ice has been completed and the pressure of the main chamber has returned to base levels using the equation:

$$N = \ln(10) \frac{S}{A_v}$$

(3)

where $S$ (cm$^{-1}$) is the integrated absorbance of a mid-infrared band characteristic to the solid-phase molecular species in question and $A_v$ is the strength constant associated with that band. It should be noted that the inclusion of the constant term $\ln(10)$ allows for the relation of $A_v$, which is measured on an optical depth scale, to $S$, which is measured on an absorbance scale.

The value of $N$ derived from acquired spectra can be related to the total thickness of the ice $d$ through the equation:

$$d = 10^4 \frac{m}{\rho N_A} N = 10^4 \ln(10) \frac{mS}{\rho N_A A_v}$$

(4)

where $m$ is the molar mass (g mol$^{-1}$) of the molecular species in question, $\rho$ (g cm$^{-3}$) is the temperature-dependent solid-phase density, and $N_A$ is the Avogadro constant (6.02×10$^{23}$ molecules mol$^{-1}$). Note that, in this case, the constant 10$^4$ allows for $d$ to be expressed in units of μm.

In the case of multi-component ices, a first approximation of the ice thickness may be obtained by applying Eq. (4) to each component of the ice mixture and then summing up the contributions of the individual molecular species to the total thickness. However, it should be noted that, although Eq. (4) may provide for a relatively straightforward means of assessing the thickness of a multi-component ice from acquired mid-infrared absorption spectra, it is likely that the overall thickness value derived from such an approach is associated with rather large uncertainties. These uncertainties mainly derive from the use of literature values of $A_v$ which are typically quoted for pure molecular ices but which may vary by as much as 50% for molecular components of a solid-phase mixture **[39]**.

It should also be noted that, in the case of ices prepared by background deposition, the total ice observed using mid-infrared spectroscopy is the summation of the ice deposition on the front and rear sides of the deposition substrate (Figures 4 and 5), only the former of which is exposed to processing by ion beams. It is therefore desirable to minimise the amount of ice deposited on the rear side of the substrate which is not processed radiolytically but which is nonetheless observed by the mid-infrared spectroscopic beam. The sample holder presents a cylindrical hole of diameter 10 mm and length 17 mm on its rear side which permits the mid-infrared spectroscopic beam to be transmitted through the deposited ice to the external detector, but which also ensures minimal condensation of material onto the rear side of the deposition substrate. Using this configuration, we calculate that the total column density of ice deposited onto the rear side of the deposition substrate during background deposition is less than 10% of the total ice deposited on the front side (see our previous work published in this journal **[27]** for more information on how this value is calculated).

Up until this point in the discussion, the preparation of an astrochemical ice analogue required that its precursor material is sufficiently volatile such that it is a gas at standard temperature and pressure or else that it produces vapours that may be readily introduced into the pre-dosing mixing container. It is, however, also possible to prepare ices from more refractory materials such as, for example, polycyclic aromatic hydrocarbons. This may be achieved by installing a removable effusive evaporator (Createc OLED-40-10-WK-SHM) onto one of the side flanges of the AQUILA chamber; typically in place of the laser used for the interferometric determination of ice thickness growth rates. This effusive evaporator is pre-loaded with a quartz tube containing a sample of the refractory material which may be heated to temperatures of up to 800 °C under UHV conditions, thereby allowing sublimation of the solid material and its introduction into the main chamber. To minimise the possibility of the refractory material contaminating the chamber by coating any internal surface other than the deposition substrate, the sample holder and deposition substrate are rotated to directly face the effusive evaporator during the deposition of the refractory material, and the nozzle of the effusive evaporator is brought into close proximity of the deposition substrate by means of a manually operated bellows. Due to the need to rotate the sample holder during the deposition of a refractory material, it is not possible to monitor ice growth *in situ*. Instead, an ice must be grown in progressive steps after each of which the sample holder is again rotated such that its surface is orthogonal to the mid-infrared beam thus allowing for a spectroscopic assessment of the amount of ice deposited.

Once an ice of a desired thickness and stoichiometric composition has been prepared, it may then be exposed to an ion beam supplied by the ECR ion source such that the projectile ions impact the target astrochemical ice analogue at an angle of 45° to the normal. Radiolysis-induced changes to the physical structure and chemical composition of the ice may be monitored *in situ* again using mid-infrared absorption spectroscopy, which allows for the destruction of the molecular components of the ice to be measured through the decreasing intensities of their corresponding absorption bands and the synthesis of new molecules to be identified by the appearance of new absorption features in the acquired spectra **[40,41]**. Moreover, physical changes to the ice such as the amorphisation of crystalline structures or the compaction of porous solids may also be studied through changes in the acquired mid-infrared spectra **[42,43]**.

Another physical process that is known to be induced by ion impact is the sputtering of ice material **[44]**. It is very difficult to determine whether the reduction in the column density of a target astrochemical ice analogue is due to its radiolytic dissociation or its sputtering as a result of ion impact if mid-infrared absorption spectroscopy is the only analysis technique available. As such, the AQUILA is also fitted with a residual gas analyser that acts as a quadrupole mass spectrometer (Figure 3) and which can provide quantitative information on the composition of the gas phase of the chamber. Thus, any closed-shell molecules or radical fragments that are removed from the ice either as a result of ion-induced sputtering or thermal desorption during controlled warming of the substrate can be identified.

Indeed, the AQUILA set-up is also suitable for performing TPD experiments, in which an astrochemical ice analogue that may or may not have been irradiated by an ion beam is warmed at a constant rate (typically 1 or 2 K min$^{-1}$). This warming allows for the spectroscopic and spectrometric study of a number of thermally induced processes that are directly relevant to interstellar physics and chemistry, such as the diffusion of radicals within the ice matrix, the crystallisation of the ice, thermal desorption of molecules and radicals, sublimation dynamics, and thermally induced chemical reactions at low temperatures **[28,45-48]**.

## 4. Ion Irradiation of Condensed CH$_3$OH

To illustrate the utility of the AQUILA set-up in studying the radiation chemistry in astrochemical ice analogues induced by projectile ions, we have performed four irradiations of amorphous CH$_3$OH ice at 20 K using 5 keV H$^+$, 10 keV H$^+$, 10 keV O$^+$, and 20 keV O$^{2+}$ ions. Our choice of CH$_3$OH as the target ice is based not only on the fact that this molecule is one of the most ubiquitous in interstellar space **[14]**, but also on the rich chemistry that is known to occur when this ice is subjected to ionising radiation which has been reported to result in the formation of complex organic molecules, among other species **[49-57]**. As such, CH$_3$OH represents an ideal test material for showcasing the experimental capabilities of the AQUILA set-up.

Amorphous CH$_3$OH ices having an approximate thickness of 300 nm were prepared by background deposition of the vapour at 20 K and a pre-irradiation mid-infrared absorption spectrum was acquired (Figure 6). Each ice was subsequently exposed to one of the aforementioned ion beams to simulate the processing of an interstellar CH$_3$OH ice by cosmic ionising radiation, and additional spectra were collected at pre-determined ion fluence intervals. In each case, absorption bands attributable to CH$_3$OH were noted to decrease in intensity as the experiment progressed, while new bands emerged due to the radiolytic formation of product molecules of which the most pertinent to this article are CO, CO$_2$, and CH$_4$ (Figure 7). The extent of CH$_3$OH destruction and molecular product formation could be quantified through the measurement of their respective column densities by using Eq. (3), as described previously. Table 1 lists the absorption bands and the associated integrated band strength constants used to quantify the column density of each of these molecules.

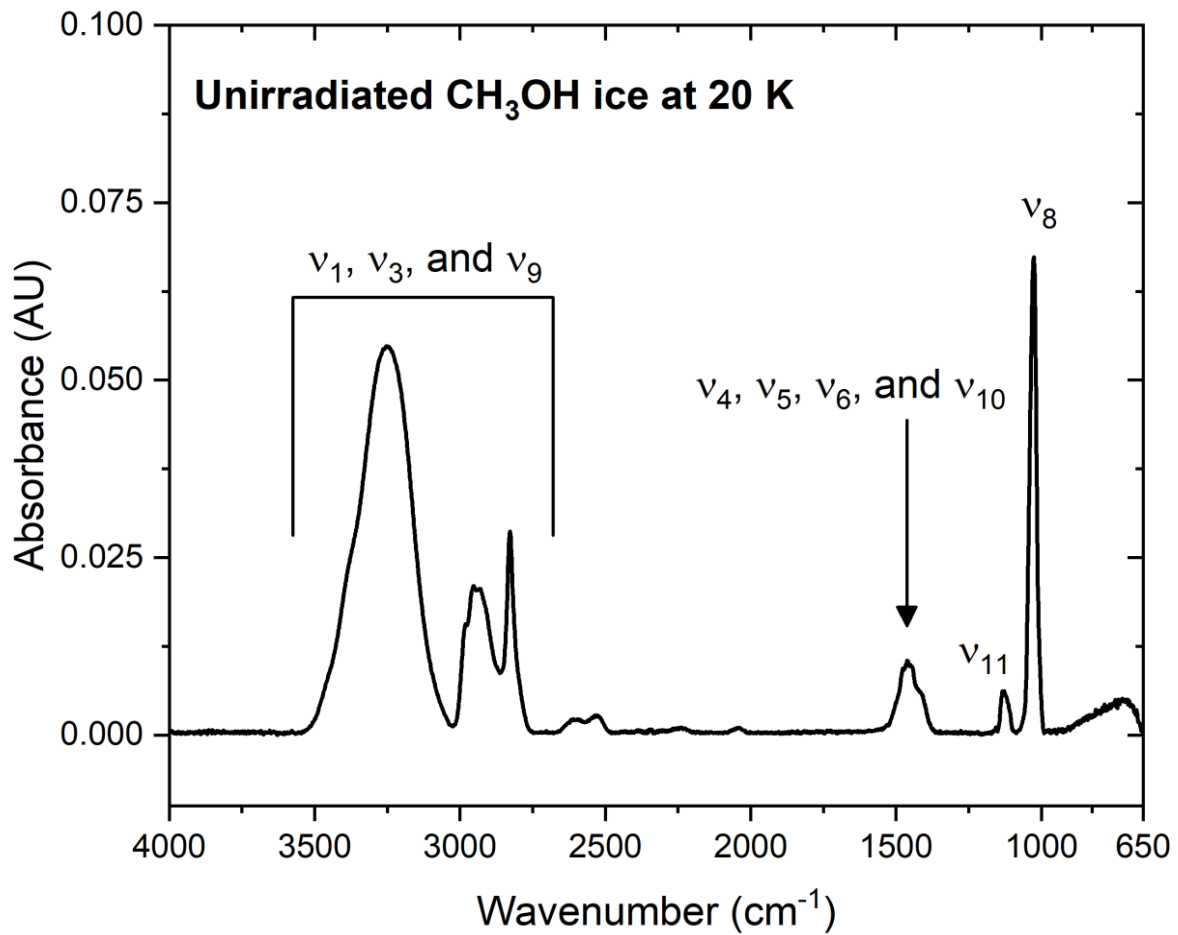

**Figure 6.** Mid-infrared absorption spectrum of a pristine, unirradiated CH$_3$OH ice at 20 K. The major vibrational modes are indicated and are based on assignments made by Gálvez *et al.* **[58]** and Bouilloud *et al.* **[59]**.

**Table 1.** Positions of the mid-infrared absorption bands ($v$) and their associated band strength constants ($A_v$) used to quantify the column densities of different molecules from acquired spectra.

| Molecule | $v$ (cm$^{-1}$) | $A_v$ (10$^{-17}$ cm molecule$^{-1}$) | Reference |
|---|---|---|---|
| CH$_3$OH | 1027 ($v_8$) | 1.61 | Luna *et al.* **[60]** |
| CO | 2138 ($v_s$) | 0.87 | González-Díaz *et al.* **[61]** |
| CO$_2$ | 2329 ($v_3$) | 11.80 | Gerakines and Hudson **[62]** |
| CH$_4$ | 1300 ($v_4$) | 1.04 | Gerakines and Hudson **[63]** |

The radiolytic destruction of CH$_3$OH ice with increasing ion fluence is depicted in Figure 8, which demonstrates that the radiolytic decay of this molecule follows an exponential-like decay trend for all projectile ions at 20 K. Indeed, we have found that the column density trend (normalised to the initial column density of CH$_3$OH) is best fitted by the sum of two exponential decay functions of the type $N = a_1\exp(-F/b_1) + a_2\exp(-F/b_2) + c$, where $a$, $b$, and $c$ are constants and $F$ is the delivered ion fluence in units of ions cm$^{-2}$. It is possible to note significant differences in the rates and extent of CH$_3$OH destruction induced by the different ion beams. For example, although the irradiation of CH$_3$OH ice by 10 keV H$^+$ ions results in a more rapid destruction that does the analogous irradiation by 5 keV H$^+$ ions (i.e., a faster radiolytic destruction is observed when using higher energy H$^+$ ions), the same is not true for irradiations using 10 keV O$^+$ and 20 keV O$^{2+}$ ions, wherein the former results in a more rapid and extensive destruction of CH$_3$OH than the latter.

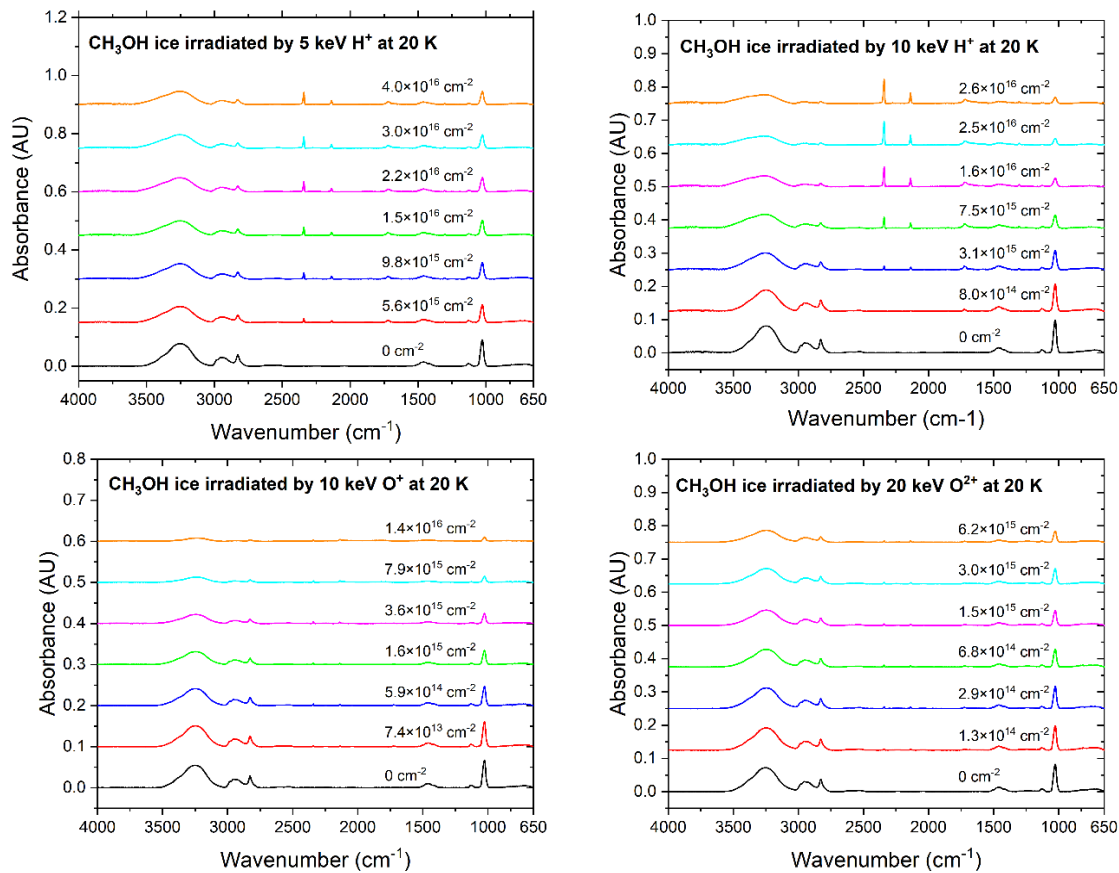

**Figure 7.** Mid-infrared absorption spectra of CH$_3$OH ices irradiated by different ions. Projectile ion fluences are indicated above each individual spectrum. Note that spectra are vertically shifted for clarity.

Such an observation may be rationalised by considering the dominant mechanisms of energy loss for each of the projectile ions utilised. As can be seen from the data displayed in Table 2, which were calculated using the *SRIM: Stopping and Range of Ions in Matter* programme [64], the dominant energy loss mechanism for the H$^+$ beams is electronic (or inelastic) stopping while for the oxygen ion beams it is nuclear (or elastic) stopping. Electronic stopping is generally expected to be the dominant method of energy loss for a projectile ion with a kinetic energy greater than 1 keV amu$^{-1}$. Thus, for the 10 keV O$^+$ ions used in this study, nuclear stopping is the dominant energy loss mechanism. Although the overall contribution of nuclear stopping to ion energy loss is lower in the case of the 20 keV O$^{2+}$ ions, this decreasing contribution occurs at a significantly slower rate than the increasing contribution of the electronic stopping, thereby explaining the smaller differences in the CH$_3$OH destruction trends induced by 10 and 20 keV oxygen ion irradiation compared to the relatively larger differences induced by 5 and 10 keV H$^+$ ion irradiation.

As mentioned previously, the radiolytic destruction of CH$_3$OH results in the formation of a number of product molecules, the most pertinent of which for the purposes of this discussion are CO, CO$_2$, and CH$_4$. A number of radiolytic reaction pathways that lead to the formation of these molecules have been identified, many of which were described in great detail in our previous works **[27,28,65]** and in those of other research groups **[50,53,56]**. A summary of the reaction network leading to these product molecules is given in Figure 9.

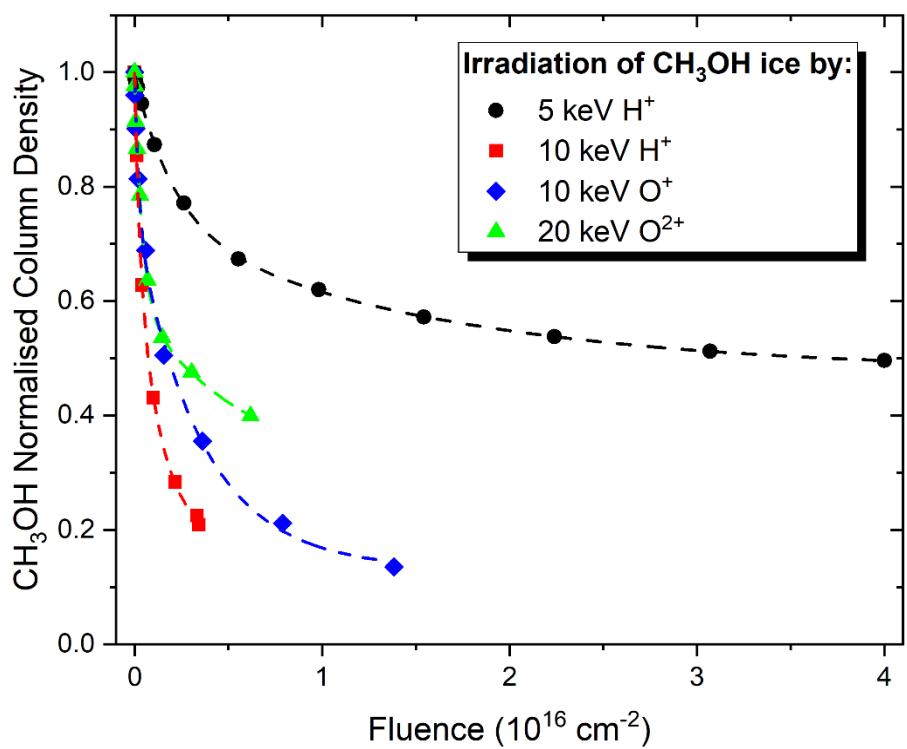

**Figure 8.** Column density of CH$_3$OH normalised to its initial column density prior to irradiation as a function of projectile ion fluence. The dashed lines correspond to the sum of two exponential decay curves fitted to the data (more detailed information may be found in the text).

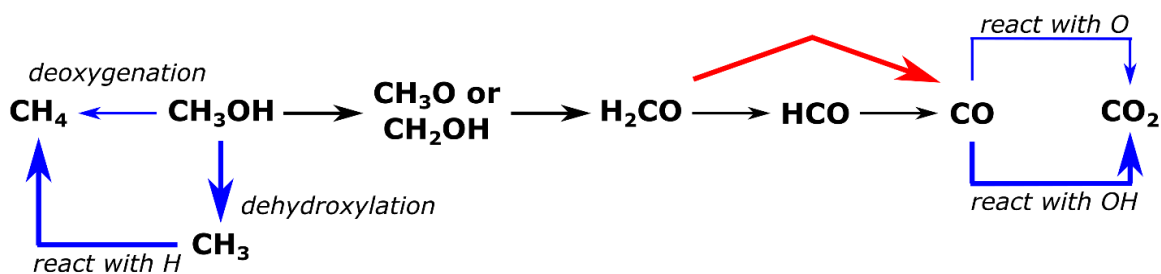

**Figure 9.** Reaction network for the synthesis of CO, CO$_2$, and CH$_4$ in an irradiated CH$_3$OH ice, including all intermediate radicals and molecules. Dehydrogenation reactions are symbolised by black arrows, one-step double dehydrogenations are symbolised by red arrows, and other types of reactions are symbolised by blue arrows. Note that, in the case of the existence of multiple pathways towards a common molecule, the most favourable pathway (as determined by the previous works of Bennett *et al.* **[50]**, Schmidt *et al.* **[56]**, and Ioppolo *et al.* **[66]**) is indicated by a bold typeset arrow.

Table 2. Nuclear, electronic, and total stopping, as well as penetration depths, of ions used in this study as calculated using the SRIM programme [64]. Note that SRIM is only able to consider singly charged ions, and so the stopping for 20 keV $O^{2+}$ ions is approximated by that for 20 keV $O^+$ ions.

| Projectile Ion | Stopping (eV nm$^{-1}$) | | | Penetration Depth (nm) |
| --- | --- | --- | --- | --- |
| | Nuclear | Electronic | Total | |
| 5 keV $H^+$ | 0.95 | 23.70 | 24.65 | 152 |
| 10 keV $H^+$ | 0.59 | 32.73 | 33.32 | 258 |
| 10 keV $O^+$ | 98.67 | 47.26 | 145.93 | 48 |
| 20 keV $O^{2+}$ | 82.77 | 66.83 | 149.60 | 91 |

It is at this point that we turn our attention to the abundances of CO, $CO_2$, and $CH_4$ produced in our irradiation experiments. These may be analysed in two ways: (i) the amount of a product that is yielded when $CH_3OH$ ice is processed by different ion beams, and (ii) the ratio of one product to another that is yielded after irradiation using each beam. These data are summarised in Figure 10, which depicts the molecular column densities of CO, $CO_2$, and $CH_4$ normalised to the initial column density of $CH_3OH$ throughout each irradiation in its left-hand side panels, and the ratios of these product molecules to one another throughout each irradiation in its right-hand side panels.

By examining Figure 10, it is possible to note that the formation rates of each of the three products under consideration behave somewhat differently to one another. In the case of CO, for instance, it is possible to note that irradiation using each of the four beams that were investigated resulted in an initial rapid increase in the column density of this product relative to the initial column density of $CH_3OH$ (i.e., the normalised column density of CO). For the 5 keV $H^+$ beam and the 10 keV $O^+$ beam, a similar trend was apparent in which the normalised column density of CO initially increased rapidly before plateauing. In the case of the 10 keV $H^+$ beam, it is evident that the plateau had not yet been reached by the end of irradiation, and that the normalised column density of CO was still on the rise. The case of the 20 keV $O^{2+}$ ion beam provides for an interesting study, as it is evident that the normalised column density of CO initially increased rapidly before plateauing for a short fluence interval of about $1.2 \times 10^{15}$ ions cm$^{-2}$ and subsequently decreasing slightly. It is possible that, had longer fluences been used for the other ion beams, then similar trends with respect to the normalised column density of CO would have been observed in those cases.

The trends observed for the normalised column density of $CO_2$ are very similar to those of CO (Figure 10). This is perhaps to be expected, since CO is the direct precursor to $CO_2$ in the reaction network shown in Figure 9, and thus the abundance of the former should correlate with that of the latter. One notable exception, however, is the formation trend of $CO_2$ during the irradiation of $CH_3OH$ by 10 keV $O^+$ ions, where the normalised column density of the product initially increases rapidly, plateaus briefly, and then declines slightly. A number of physical and chemical factors may contribute to this discrepancy between the CO and $CO_2$ formation profiles during irradiation by 10 keV $O^+$ ions, including sputtering of the molecular components of the ice (including CO and $CO_2$ themselves), fewer interactions with the electrons of target molecules due to the dominance of nuclear stopping, or even possible perturbation of the chemical reaction network induced by the implantation of the oxygen ion.

The formation trends of $CH_4$ as a result of the irradiative processing of $CH_3OH$ are broadly similar across all investigated ion beams. In general, it is possible to note that the normalised column density of $CH_4$ initially increases very rapidly and peaks at ion fluences significantly lower than for CO and $CO_2$, before subsequently decaying rapidly as the nascent $CH_4$ is destroyed by ion irradiation at higher fluences. Such a result agrees well with previously published reports on the ion irradiation of $CH_3OH$ [27,67-69], which have also demonstrated $CH_4$ to be a product that is quickly formed and destroyed. It is to be noted that, although a similar formation / destruction trend is noted for the $CH_4$ yielded as a result of the 10 keV $O^+$ ion irradiation of $CH_3OH$, this is skewed towards significantly higher fluences than for the other ion beams considered in this study. It is to be recalled that the dominant mechanism of energy loss for 10 keV O$^+$ ions is nuclear stopping (Table 2), unlike for the other ion beams where electronic stopping is either dominant (in the case of the 5 and 10 keV H$^+$ ions) or a large contributor (in the case of the 20 keV O$^{2+}$ ions) to energy loss. This therefore explains the somewhat different behaviour observed with regards to the radiolytic formation / destruction of CH$_4$ when using 10 keV O$^+$ ions.

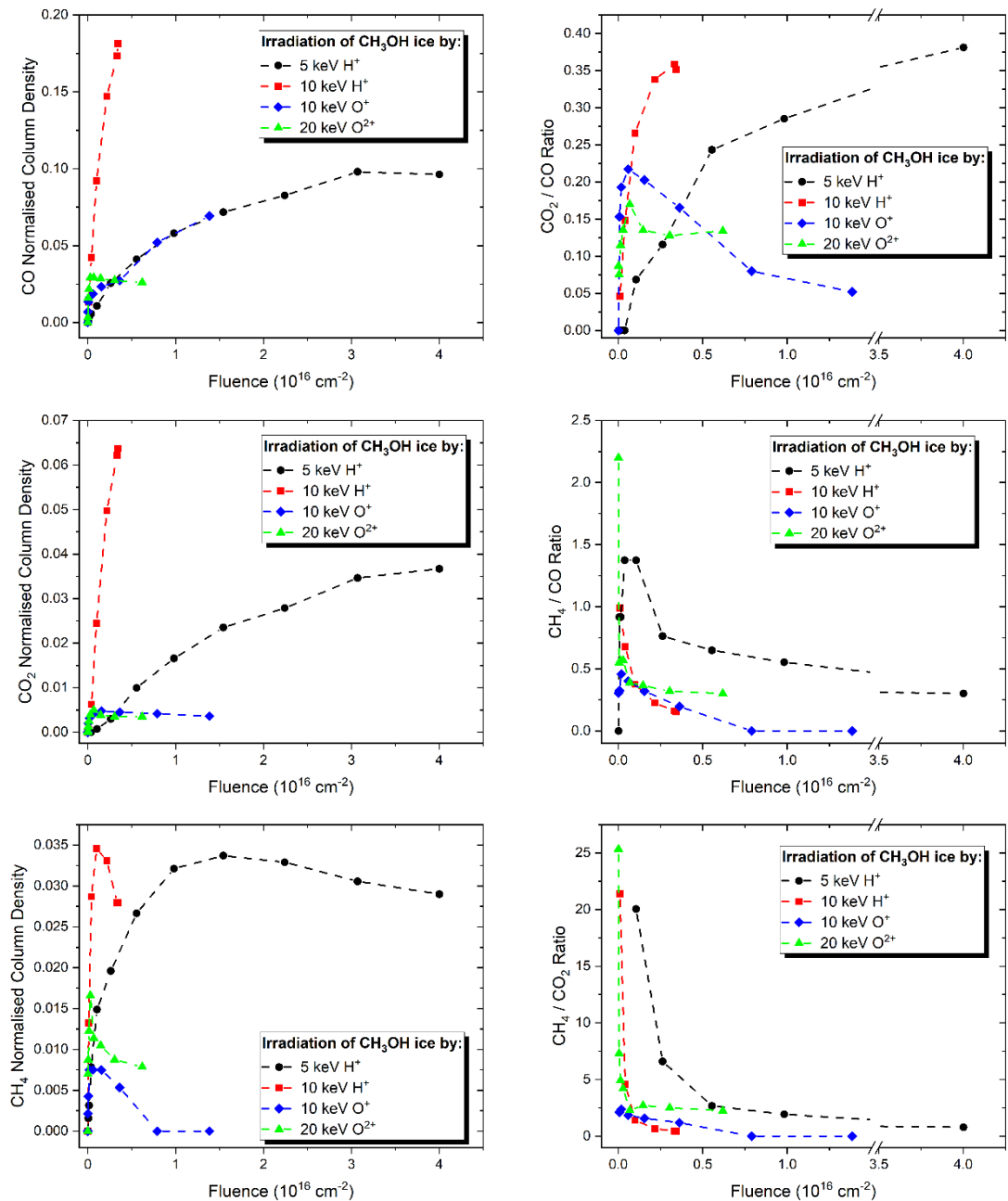

**Figure 10.** *Left panels:* Molecular column densities of CO (top), CO$_2$ (middle), and CH$_4$ (bottom) normalised to the initial column density of CH$_3$OH as measured during each irradiation. *Right panels:* Plots showing the CO$_2$ / CO (top), CH$_4$ / CO (middle), and CH$_4$ / CO$_2$ (bottom) ratios as measured during each irradiation. Note that the dashed lines in the panels are not fits and are potted solely to guide the eye.

Finally, it is worth briefly explaining the trends observed with regards to the abundance of each of the three product molecules under consideration relative to one another, as shown in the right-hand panels of Figure 10. In the case of the $CO_2$ / CO ratio, it is possible to note that, for all ion beams investigated, this ratio consistently remained below unity during irradiation and has a qualitatively similar trend to that of the $CO_2$ normalised column density evolution. Such results are expected given that the formation of $CO_2$ within the irradiated $CH_3OH$ ice is dependent upon the reaction of CO.

The calculated ratios of $CH_4$ to CO and $CO_2$ are also interesting, as they confirm that the production of $CH_4$ in the irradiated $CH_3OH$ ice is rapid; indeed, in some irradiations, large abundances of $CH_4$ are observed before any CO or $CO_2$ becomes quantifiable thus resulting the first plotted data point in the middle- and bottom-right panels of Figure 10 being well in excess of unity. In the case of the 20 keV $O^{2+}$ ion beam, the abundance of $CH_4$ is initially about 2.25 and 25 times greater than that of CO and $CO_2$, respectively, when the abundances of these latter molecules first become quantifiable. However, as each irradiation progresses, the propensity of $CH_4$ to undergo radiolytic destruction significantly reduces the $CH_4$ / CO and $CH_4$ / $CO_2$ ratios. All $CH_4$ / CO and $CH_4$ / $CO_2$ ratios, with the exception of the $CH_4$ / $CO_2$ ratio obtained at the end of the 20 keV $O^{2+}$ ion beam irradiation, were reduced to less than unity by the end of the experiment thereby indicating that $CH_4$ remained as a minor product in comparison to CO and $CO_2$. It should be noted, however, that although the $CH_4$ / $CO_2$ ratio was not reduced to less than unity by the end of the 20 keV $O^{2+}$ ion beam irradiation experiment, it did decrease tenfold.

## 5. Temperature-Programmed Desorption of Pristine and Ion Irradiated $CH_3OH$

To further illustrate the utility of the AQUILA set-up in studying processes relevant to astrochemistry and laboratory astrophysics, we have performed TPD studies of $CH_3OH$ ice, both pristine and irradiated using 10 keV $H^+$ ions at 20 K. As mentioned previously, thermally induced processes play important roles in the evolution of ices in the interstellar medium and outer Solar System [28,45-48], and could contribute to the formation of molecules of significance to biochemical or geochemical processes. Our first experiment involved the background deposition of a 270 nm-thick $CH_3OH$ ice at 20 K which was subsequently warmed at a rate of 1 K min$^{-1}$ until complete sublimation. Throughout this warming process, the solid-phase ice and the gas composition of the AQUILA chamber were continuously monitored *via* mid-infrared absorption spectroscopy and quadrupole mass spectrometry, respectively.

Figure 11 depicts the integrated absorbance over the 1060-990 cm$^{-1}$ wavenumber range of the mid-infrared spectrum of $CH_3OH$, corresponding to the $v_8$ vibrational mode (Table 2). As can be seen, the integrated absorbance increases slightly with increasing temperature, from 2.12 cm$^{-1}$ at 20 K to 2.31 cm$^{-1}$ at 109 K. This seemingly contradictory trend is actually due to the sharpening and narrowing of the mid-infrared absorption bands of $CH_3OH$ as it undergoes a phase change from an amorphous solid to a crystalline structure [58,65]. As the temperature of the ice is further raised, a sharp drop in the value of the integrated absorbance is observed at a temperature of about 110 K. This phenomenon is likely due to a combination of thermally induced pore collapse coupled to the release of a limited number of $CH_3OH$ molecules as the exothermic amorphous-to-crystalline phase transition is completed [70,71]. At even higher temperatures, the integrated absorbance begins to decline gradually due to sublimation-induced losses of the solid material. This is most noticeable at temperatures greater than 140 K and, indeed, by 148 K only about 1% of the initial integrated absorbance of $CH_3OH$ can be measured in the acquired mid-infrared absorption spectrum.

Figure 11 also depicts the *m*/*q* = 31 signal measured through quadrupole mass spectrometry as a function of ice temperature. This particular *m*/*q* ratio is ascribed to either the $CH_3O^{\bullet+}$ or the $CH_2OH^{\bullet+}$ radical ion fragments that are produced within the mass spectrometer as a result of radiolysis of the parent $CH_3OH$ molecule by 70 eV electron impact. The rationale behind following this *m*/*q* ratio as opposed to that of the parent molecular ion at *m*/*q* = 32 is the known dominance of the former in the mass spectrum of pure $CH_3OH$, as noted in the NIST database. It is possible to note that the $m/q = 31$ signal remains relatively constant at background noise levels at temperatures below about 130 K. Above this temperature, the signal rises sharply due to the sublimation of solid $CH_3OH$ to the gas phase, where it peaks at a temperature of 147 K. The qualitative and quantitative agreement of the mid-infrared and mass spectrometric data depicted in Figure 11 with each other, as well as with published literature data [72], support the appropriateness of the AQUILA for performing TPD studies.

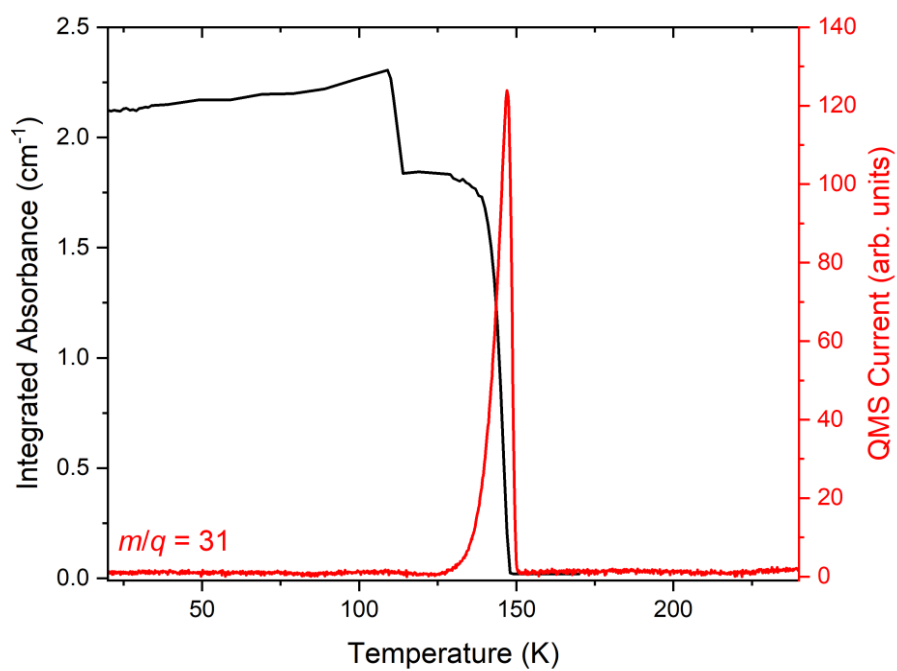

**Figure 11.** Results from the TPD experiment of pristine $CH_3OH$ ice deposited at 20 K and subsequently warmed at 1 K min$^{-1}$. The black curve represents the variation of the integrated absorbance of the $\nu_8$ mode of $CH_3OH$ ice as measured through mid-infrared absorption spectroscopy as a function of temperature, while the red curve represents the $m/q = 31$ signal recorded by the quadrupole mass spectrometer corresponding to the $CH_3O^{\bullet+}$ or the $CH_2OH^{\bullet+}$ radical ion fragments in the gas phase.

Our second experiment involved the background deposition of an identical 270 nm-thick $CH_3OH$ ice at 20 K which was irradiated using 10 keV $H^+$ ions to a final accumulated fluence of $5\times10^{15}$ ions cm$^{-2}$, after which the irradiated ice was subjected to TPD at a rate of 1 K min$^{-1}$. Figure 12 depicts the $m/q = 31$ signal recorded by the quadrupole mass spectrometer during the irradiation of $CH_3OH$ ice at 20 K. As can be seen, noticeable signal increases are associated with the stepwise deliverance of $H^+$ ions, which may be attributed to ion-induced sputtering of the $CH_3OH$ ice during irradiation. Such a result is particularly interesting since $H^+$ ions are acknowledged to be less efficient at inducing the sputtering of astrochemical ice analogues compared to heavier ions [25,26], although it should be noted that $H^+$ ion-induced sputtering has been reported for a number of ices, including $H_2O$ and $O_2$ [73,74].

After reaching a total fluence of $5\times10^{15}$ $H^+$ ions cm$^{-2}$, the ice was warmed at a rate of 1 K min$^{-1}$ as had been the case during the TPD experiment of unirradiated $CH_3OH$ ice. As was discussed throughout Section 4, the irradiative processing of $CH_3OH$ ice results in the formation of a number of product molecules, the most relevant to this article being CO, $CO_2$, and $CH_4$ (Figures 7 and 10). Indeed, trends in the mass spectrometric signals at $m/q = 16$, 28, and 44 (respectively corresponding to the $CH_4^{\bullet+}$, $CO^{\bullet+}$, and $CO_2^{\bullet+}$ radical ion fragments), together with that at $m/q = 31$, could be measured during the TPD experiment, as shown in Figure 13.

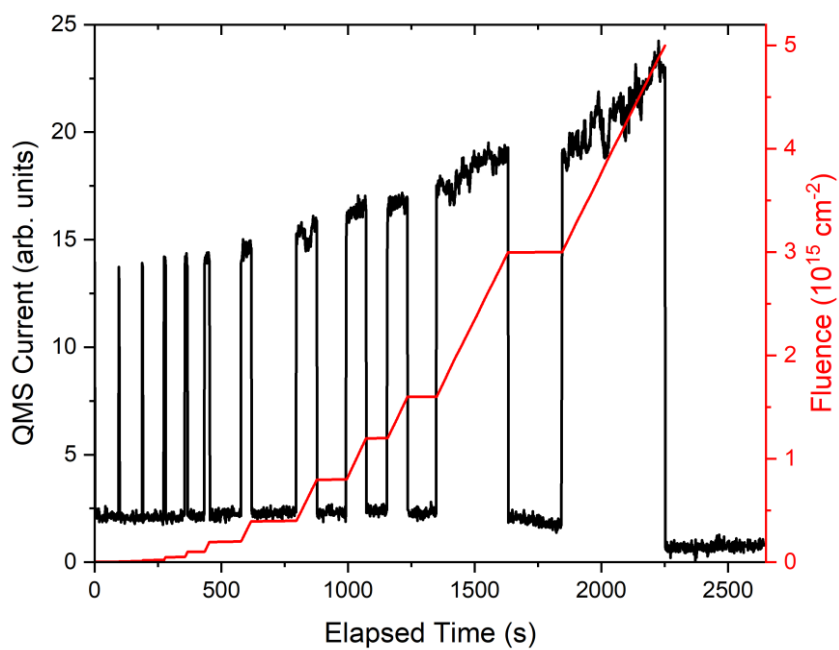

**Figure 12.** The $m/q = 31$ mass signal attributed to the $CH_3O^{•+}$ or the $CH_2OH^{•+}$ gas-phase radical ion fragments arising from the sputtering of $CH_3OH$ ice is correlated to the delivery of 10 keV $H^+$ ions during the stepwise irradiation of the ice at 20 K.

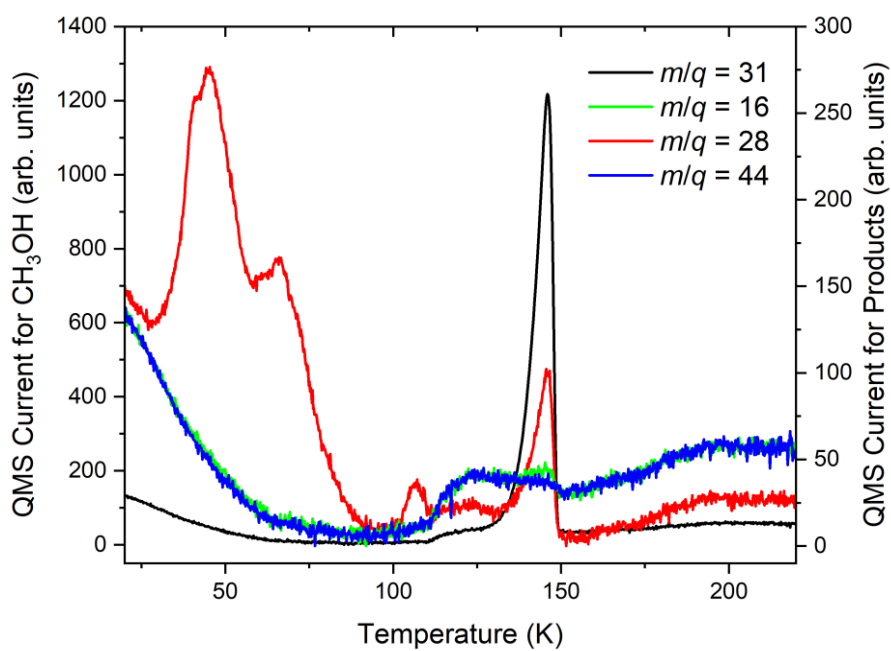

**Figure 13.** TPD experiment of a $CH_3OH$ ice irradiated by 10 keV $H^+$ ions at 20 K. Shown in the figure are the $m/q = 31$ ($CH_3O^{•+}$ or the $CH_2OH^{•+}$ radical ion fragments; black curve), 16 ($CH_4^{•+}$ radical ions; green curve), 28 ($CO^{•+}$ radical ions; red curve), and 44 ($CO_2^{•+}$ radical ions; blue curve) mass signals as a function of temperature. Note that the *y*-axis on the left-hand side is associated with the data for $m/q = 31$, while the right-hand side *y*-axis is associated with all other data.

It is interesting to note that the measured quadrupole mass spectrometric signals for desorbing $CH_3OH$ displayed in Figures 11 and 13 are similar to one another, except for a slight enhancement in the desorption of $CH_3OH$ over the 111-128 K temperature range in the ion irradiated ice. This temperature range is associated with the maximum rate of thermally induced crystallisation of amorphous $CH_3OH$ **[58,65]**, and thus it is conceivable that some $CH_3OH$ molecules were lost from the ice to the gas phase during this exothermic process. The observed enhancement may have arisen due to the fact that the ion irradiated ice was likely more disordered than its unirradiated counterpart, due to the structural defects introduced by ion irradiation and sputtering during 10 keV $H^+$ ion irradiation. Nevertheless, it is to be noted that the maximum desorption of $CH_3OH$ from the ion irradiated ice occurred at 146 K (Figure 13), which is within 1 K of the analogous result reported for the unirradiated $CH_3OH$ ice (Figure 11).

The radiolytic products CO, $CO_2$, and $CH_4$ also desorbed from the ice during the TPD experiment. In the case of CO, a prolonged desorption starting at 27 K and peaking at 45 K was recorded at temperatures below 100 K (Figure 13). This peak desorption temperature of 45 K is somewhat higher than that recorded during the TPD of pure CO ice, likely due to the trapping of the volatile CO molecules by less volatile ice components **[75,76]**; primarily $CH_3OH$. CO desorption begins to peak again at a temperature of 107 K, just as the lower abundance $CO_2$ and $CH_4$ components of the ice also begin to desorb (Figure 13). In this scenario, as $CO_2$ and $CH_4$ molecules thermally desorb from the ice, they create sufficient channels and defects in the ice structure that permit the trapped CO molecules to also escape into the gas phase. Indeed, this effect is even more pronounced at 146 K, when any remnant CO molecules desorb into the gas phase as the $CH_3OH$ ice undergoes bulk sublimation.

## 6. Comparison with the ICA Apparatus

As a final discussion point, we wish to contextualise our description of the AQUILA set-up and its capabilities by comparing it with another laboratory astrochemistry facility at Atomki: the ICA **[27,28]**. A comparative summary of the two set-ups is provided in Table 3. The basic design and operational function of both chambers is essentially the same, in the sense that both are UHV chambers operating at base pressures of $10^{-9}$ mbar and both contain an oxygen-free high-conductivity copper sample holder into which one or more infrared-transparent deposition substrates may be mounted and cooled to a temperature of between 20 and 300 K with the use of closed-cycle helium cryostats. Astrochemical ice analogues may be prepared on these infrared-transparent substrates *via* either background or direct deposition of dosed gases and vapours, and subsequently processed by ion beams. The physical and chemical changes induced by this irradiation may be monitored by the same analytical techniques at both set-ups, namely Fourier-transform mid-infrared transmission absorption spectroscopy and quadrupole mass spectrometry.

Despite these operational similarities, a number of technical differences exist between the AQUILA and ICA set-ups. Perhaps most importantly, the ion beams delivered by the ECR ion source to the AQUILA are of significantly lower energy (50 eV to 900 keV) than are those delivered by the Tandetron accelerator to the ICA (200 keV to 10 MeV). This therefore allows for a wide range of energies to be used when performing ion irradiation experiments at Atomki, thus allowing us to simulate the energetic processing of astrochemical ice analogues by various types of radiation relevant to astrophysics and planetary science, including lower energy ions representative of those in the solar wind or those emitted during the decay of radioactive isotopes, as well as higher energy ions more characteristic of galactic cosmic rays.

Other differences between the set-ups include the fact that the ECR ion source is able to routinely deliver multiply charged ions, molecular ions, and negatively charged ions to the AQUILA for use in radiation astrochemistry experiments, while the Tandetron accelerator is more limited in this regard in that it routinely produces singly or doubly charged atomic ions. Conversely, it is possible to combine ion irradiation with electron irradiation at the ICA by virtue of an electron gun that is affixed to that chamber, whereas irradiation using electrons as projectiles is currently not possible at the AQUILA. Moreover, the sample holder used in the ICA is capable of hosting up to four infrared-transparent deposition substrates which may be sequentially irradiated thereby allowing experiments to be repeated with ease. This is advantageous compared to the single deposition substrate that may presently be hosted by the sample holder within the AQUILA. However, it is to be noted that the presence of a single deposition substrate at the AQUILA allows for post-irradiative TPD studies to be carried out fairly easily. This is not the case at the ICA, where TPD studies can only be carried out for irradiated ices deposited onto a single deposition substrate by direct deposition, thereby excluding the thermally induced sublimation of material from other deposition substrates.

It is important to note that, despite their differences, neither the AQUILA nor the ICA should be considered to be the successor of the other; rather, these experimental set-ups operate in a complementary manner that allows us to fully investigate a range of radiation-induced physical and chemical processes that are relevant to astrochemistry. In this way, the Atomki astrochemistry laboratories present a unique opportunity for experimentalists to probe practically every aspect of particle-induced radiation astrochemistry at a single large-scale facility, thus allowing for the scope and scale of various research projects to be expanded without significant logistical or operational difficulties.

**Table 3.** Comparisons of the major features of the AQUILA and ICA laboratory astrochemistry set-ups at Atomki. Note that information on the user demand on the ECR ion source and Tandetron accelerator is the average over the years 2020-2022.

| Parameter or Feature | AQUILA | ICA |
|---|---|---|
| *Accelerator Features* | | |
| Associated ion source | ECR ion source | Tandetron accelerator |
| Element range of projectile ions | H to Au | H to Au |
| Energy range of projectile ions | 50 eV to 900 keV | 200 keV to 10 MeV |
| Charge-state of projectile ions | Singly to multiply | Singly or doubly |
| Nominal current densities used | Up to a few μA | A few hundred nA |
| Availability of negatively charged ions | Yes | No |
| Production of molecular ions | Routine | Infrequent |
| User demand on ion source (hours per year) | 180 | 2469 |
| | | |
| *Chamber Facilities* | | |
| Chamber inner diameter (mm) | 300 | 160 |
| Operational base pressure (mbar) | $10^{-9}$ | $10^{-9}$ |
| Operational temperature range (K) | 20-300 | 20-300 |
| Sample holder material | OFHC Cu | Gold-coated OFHC Cu |
| Deposition substrates used simultaneously | 1 | Up to 4 |
| Availability of electron irradiation | No | Yes |
| Deposition method for gases and vapours | Direct and background | Direct and background |
| Deposition method for solids | Effusive evaporation | Effusive evaporation |
| Availability of *in situ* spectroscopic methods | FTIR absorption | FTIR absorption |
| Availability of *in situ* spectrometric methods | QMS | QMS |
| Availability of laser interferometric methods | Yes | No |

# 7. Conclusions

In this article, we have described in detail a new laboratory astrochemistry apparatus named the AQUILA, which has been installed as a permanent end-station to the ECR ion source at the HUN-REN Institute for Nuclear Research (Atomki) in Debrecen, Hungary. This set-up allows for the keV ion irradiation of astrochemical ice analogues representative of those that exist on the surfaces of icy outer

Solar System bodies or adsorbed on interstellar dust grains to be studied. Processes induced by cosmic radiation such as that simulated by the ion beam supplied by the Atomki ECR ion source may play key roles in the chemical evolution of astronomical environments by driving the synthesis of complex organic molecules, some of which may be directly relevant to the emergence of life and the development of planetary systems.

Aside from a description of the key features, operating principles, and experimental protocols of the AQUILA, we have also demonstrated the experimental capabilities of the apparatus by performing the irradiation of amorphous $CH_3OH$ at 20 K using 5 and 10 keV $H^+$ ions, as well as 10 keV $O^+$ and 20 keV $O^{2+}$ ions. Using Fourier-transform mid-infrared transmission absorption spectroscopy, we have been able to quantify the decay in the molecular column density of $CH_3OH$ during each irradiation, as well as the synthesis of the radiolytic product molecules CO, $CO_2$, and $CH_4$. Our quantitative analysis has shown that distinct trends that are likely linked to whether nuclear or electronic stopping is the dominant mechanism of energy loss of the projectile ion are evident in the production and destruction rates of all these molecules. Moreover, we have demonstrated the suitability of the AQUILA for quantitative TPD experiments using case-studies of pristine and 10 keV $H^+$ ion irradiated $CH_3OH$ ices.

Finally, we have compared the AQUILA to another laboratory astrochemistry apparatus based at Atomki; the ICA. Although both set-ups rely on the same general principles of operation and utilise similar experimental protocols, the ICA is better suited to studying radiation-driven processes induced by electrons and higher energy (i.e., a few hundred keV to a few MeV) ions, whereas the AQUILA is more appropriate for studying processes induced by lower energy (i.e., a few to a few hundred keV) ions, molecular ions, and negatively charged ions, as well as performing quantitative TPD studies. Together, the AQUILA and the ICA allow for a wide range of complementary radiation astrochemistry studies to be performed at Atomki which simulate the processing of astrochemical ice analogues by a number of radiation sources relevant to astrophysics and planetary science, including galactic cosmic rays, stellar winds, giant planetary magnetospheres, and the decay of radioactive isotopes.

**Acknowledgements**

The authors gratefully acknowledge funding from the Europlanet 2024 RI which has been funded by the European Union Horizon 2020 Research Innovation Programme under grant agreement No. 871149. This article is also based on work from the COST Action CA20129 MultIChem, supported by COST (European Cooperation in Science and Technology). Zoltán Juhász acknowledges support from the Hungarian Academy of Sciences through the János Bolyai Research Scholarship. Sergio Ioppolo is grateful to the Danish National Research Foundation for financial support through the Centre of Excellence 'InterCat' (grant agreement No. DNRF150). The authors would also like to thank Alexei V. Ivlev (Max Planck Institute for Extraterrestrial Physics, Germany) for the fruitful conversations that improved this article.

**Data Availability Statement**

The data that support the findings of this study are available from the corresponding authors upon reasonable request.

**Declaration of Interests Statement**

The authors hereby declare that this work was performed in the absence of any conflicts of interest (financial or otherwise) that may have biased the outcome of this study.

**References**

1  P Swings, and L Rosenfeld. *Astrophys. J.* **86**, 483-486 (1937).
2  T Dunham. *Publ. Astron. Soc. Pac.* **49**, 26-28 (1937).

3   A McKellar. *Publ. Astron. Soc. Pac.* **52**, 187-192 (1940).

4   A McKellar. *Publ. Astron. Soc. Pac.* **52**, 312-318 (1940).

5   WS Adams. *Astrophys. J.* **93**, 11 (1941).

6   BA McGuire. *Astrophys. J. Suppl. Ser.* **259**, 30 (2022).

7   R Rubin, G Swenson, R Benson, H Tigelaar, and W Flygare. *Astrophys. J.* **169**, L39 (1971).

8   A Belloche, R Garrod, H Müller, K Menten, I Medvedev, J Thomas, and Z Kisiel. *Astron. Astrophys.* **628**, A10 (2019).

9   R Wilson, A Penzias, K Jefferts, M Kutner, and P Thaddeus. *Astrophys. J.* **167**, L97 (1971).

10  E Herbst. *Phys. Chem. Chem. Phys.* **16**, 3344-3359 (2014).

11  CR Arumainayagam, RT Garrod, MC Boyer, AK Hay, ST Bao, JS Campbell, J Wang, CM Nowak, MR Arumainayagam, and PJ Hodge. *Chem. Soc. Rev.* **48**, 2293-2314 (2019).

12  EF van Dishoeck. *Faraday Discuss.* **168**, 9-47 (2014).

13  H Linnartz, S Ioppolo, and G Fedoseev. *Int. Rev. Phys. Chem.* **34**, 205-237 (2015).

14  KI Öberg. *Chem. Rev.* **116**, 9631-9663 (2016).

15  RL Hudson, MH Moore, JP Dworkin, MP Martin, and ZD Pozun. *Astrobiology* **8**, 771-779 (2008).

16  P de Marcellus, C Meinert, I Myrgorodska, L Nahon, T Buhse, LLS d'Hendecourt, UJ Meierhenrich. *Proc. Nat. Acad. Sci. USA* **112**, 965-970 (2015).

17  W Hagen, L Allamandola, and J Greenberg. *Astrophys. Space Sci.* **65**, 215-240.

18  BM Jones, and RI Kaiser. *J. Phys. Chem. Lett.* **4**, 1965-1971 (2013).

19  G Bazsó, IP Csonka, S Góbi, and G Tarczay. *Rev. Sci. Instrum.* **92**, 124104 (2021).

20  JA Noble, HM Cuppen, S Coussan, B Redlich, S Ioppolo. *J. Phys. Chem. C* **124**, 20864-20873 (2020).

21  B Augé, E Dartois, C Engrand, J Duprat, M Godard, L Delauche, N Bardin, C Mejía, R Martinez, G Muniz, *et al. Astron. Astrophys.* **592**, A99 (2016).

22  B Augé, T Been, P Boduch, M Chabot, E Dartois, T Madi, J Ramillon, F Ropars, H Rothard, and P Voivenel. *Rev. Sci. Instrum.* **89**, 075105 (2018).

23  S Ioppolo, G Fedoseev, T Lamberts, C Romanzin, and H Linnartz. *Rev. Sci. Instrum.* **84**, 073112 (2013).

24  S Ioppolo, Z Kaňuchová, R James, A Dawes, N Jones, S Hoffmann, NJ Mason, and G Strazzulla. *Astron. Astrophys.* **641**, A154 (2020).

25  E Muntean, P Lacerda, TA Field, A Fitzsimmons, AC Hunniford, and RW McCullough. *Surf. Sci.* **641**, 204-209 (2016).

26  E Muntean, P Lacerda, TA Field, A Fitzsimmons, WC Fraser, AC Hunniford, and RW McCullough. *Mon. Not. R. Astron. Soc.* **462**, 3361-3367 (2016).

27  P Herczku, DV Mifsud, S Ioppolo, Z Juhász, Z Kaňuchová, STS Kovács, A Traspas Muiña, PA Hailey, I Rajta, I Vajda, *et al. Rev. Sci. Instrum.* **92**, 084501 (2021).

28  DV Mifsud, Z Juhász, P Herczku, STS Kovács, S Ioppolo, Z Kaňuchová, M Czentye, PA Hailey, A Traspas Muiña, NJ Mason, *et al. Eur. Phys. J. D* **75**, 182 (2021).

29  S Biri, I Vajda, P Hajdu, R Rácz, A Csik, Z Kormány, Z Perduk, F Kocsis, and I Rajta. *Eur. Phys. J. Plus* **136**, 247 (2021).

30  S Biri, R Rácz, and J Pálinkás. *Rev. Sci. Instrum.* **83**, 02A341 (2012).

31  R Rácz, S Biri, Z Juhász, B Sulik, and J Pálinkás. *Rev. Sci. Instrum.* **83**, 02A313 (2012).

32  R Bijalwan, P Ram, and M Tiwari. *J. Phys. C* **16**, 2537 (1983).

33  M Born, and E Wolf. *Principles of Optics: Electromagnetic Theory of Propagation, Interference, and Diffraction of Light*. Elsevier (2013).

34  AM Goodman. *Appl. Opt.* **17**, 2779-2787 (1978).

35  V Kofman, J He, IL ten Kate, and H Linnartz. *Astrophys. J.* **875**, 131 (2019).

36. PA Gerakines, and RL Hudson. *Astrophys. J.* **901**, 52 (2020).

37. JW Stubbing, MR McCoustra, and WA Brown. *Phys. Chem. Chem. Phys.* **22**, 25353-25365 (2020).

38. W Rocha, M Rachid, B Olsthoorn, EF van Dishoeck, MK McClure, and H Linnartz. *Astron. Astrophys.* **668**, A63 (2022).

39. J Ding, P Boduch, A Domaracka, S Guillous, T Langlinay, X Lv, ME Palumbo, H Rothard, and G Strazzulla. *Icarus* **226**, 860-864 (2013).

40. R Baragiola. *Planet. Space Sci.* **51**, 953-961 (2003).

41. R Baragiola, M Famá, MJ Loeffler, U Raut, and J Shi. *Nucl. Instrum. Methods Phys. Res. B* **266**, 3057-3062 (2008).

42. E Dartois, B Augé, P Boduch, R Brunetto, M Chabot, A Domaracka, J Ding, O Kamalou, X Lv, H Rothard, *et al. Astron. Astrophys.* **576**, A125 (2015).

43. J Souza-Corrêa, C da Costa, and E Da Silveira. *Astrobiology* **19**, 1123-1138 (2019).

44. W Brown, and R Johnson. *Nucl. Instrum. Methods Phys. Res. B* **13**, 295-303 (1986).

45. A Miyazaki, M Tsuge, H Hidaka, Y Nakai, and N Watanabe. *Astrophys. J. Lett.* **940**, L2 (2022).

46. MS Gudipati, B Fleury, R Wagner, BL Henderson, K Altwegg, and M Rubin. *Faraday Discuss.* **245**, 467-487 (2023).

47. N Fray, and B Schmitt. *Planet. Space Sci.* **57**, 2053-2080 (2009).

48. P Theulé, F Duvernay, G Danger, F Borget, J Bossa, V Vinogradoff, F Mispelaer, and T Chiavassa. *Adv. Space Res.* **52**, 1567-1579 (2013).

49. ME Palumbo, A Castorini, and G Strazzulla. *Astron. Astrophys.* **342**, 551-562 (1999).

50. CJ Bennett, SH Chen, BJ Sun, AHH Chang, and RI Kaiser. *Astrophys. J.* **660**, 1588 (2007).

51. ALF De Barros, A Domaracka, D Andrade, P Boduch, H Rothard, and E Da Silveira. *Mon. Not. R. Astron. Soc.* **418**, 1363-1374 (2011).

52. YJ Chen, A Ciaravella, GM Muñoz Caro, C Cecchi-Pestellini, A Jiménez-Escobar, KJ Juang, and TS Yih. *Astrophys. J.* **778**, 162 (2013).

53. S Maity, RI Kaiser, and BM Jones. *Phys. Chem. Chem. Phys.* **17**, 3081-3114 (2015).

54. D Paardekooper, JB Bossa, and H Linnartz. *Astron. Astrophys.* **592**, A67 (2016).

55. N Abou Mrad, F Duvernay, T Chiavassa, and G Danger. *Mon. Not. R. Astron. Soc.* **458**, 1234-1241 (2016).

56. F Schmidt, P Swiderek, and JH Bredehöft. *ACS Earth Space Chem.* **5**, 391-408 (2021).

57. LI Tenelanda-Osorio, A Bouquet, T Javelle, O Mousis, F Duvernay, and G Danger. *Mon. Not. R. Astron. Soc.* **515**, 5009-5017 (2022).

58. Ó Gálvez, B Maté, B Martín-Llorente, VJ Herrero, and R Escribano. *J. Phys. Chem. A* **113**, 3321-3329 (2009).

59. M Bouilloud, N Fray, Y Bénilan, H Cottin, MC Gazeau, and A Jolly. *Mon. Not. R. Astron. Soc.* **451**, 2145-2160 (2015).

60. R Luna, G Molpeceres, J Ortigoso, MÁ Satorre, M Domingo, and B Maté. *Astron. Astrophys.* **617**, A116 (2018).

61. C González-Díaz, H Carrascosa, GM Muñoz Caro, MÁ Satorre, and Y Chen. *Mon. Not. R. Astron. Soc.* **517**, 5744-5755 (2022).

62. PA Gerakines, and RL Hudson. *Astrophys. J. Lett.* **808**, L40 (2015).

63. PA Gerakines, and RL Hudson. *Astrophys. J. Lett.* **805**, L20 (2015).

64. JF Ziegler, MD Ziegler, and JP Biersack. *Nucl. Instrum. Methods Phys. Res. B* **268**, 1818-1823 (2010).

65. DV Mifsud, PA Hailey, P Herczku, B Sulik, Z Juhász, STS Kovács, Z Kaňuchová, S Ioppolo, RW McCullough, B Paripás, *et al. Phys. Chem. Chem. Phys.* **24**, 10974-10984 (2022).

66. S Ioppolo, Y van Bohemeen, H Cuppen, EF van Dishoeck, and H Linnartz. *Mon. Not. R. Astron. Soc.* **413**, 2281-2287 (2011).

67. G Baratta, G Leto, and ME Palumbo. *Astron. Astrophys.* **384**, 343-349 (2002).

68 F Islam, G Baratta, and ME Palumbo. *Astron. Astrophys.* **561**, A73 (2014).

69 DV Mifsud, P Herczku, B Sulik, Z Juhász, I Vajda, I Rajta, S Ioppolo, NJ Mason, G Strazzulla, and Z Kaňuchová. *Atoms* **11**, 19 (2023).

70 K Isokoski, JB Bossa, T Triemstra, and H Linnartz. *Phys. Chem. Chem. Phys.* **16**, 3456-3465 (2014).

71 B Maté, Ó Gálvez, VJ Herrero, and R Escribano. *Astrophys. J.* **690**, 486-495 (2009).

72 F Kruczkiewicz, F Dulieu, AV Ivlev, P Caselli, BM Giuliano, C Ceccarelli, and P Theulé. *Astron. Astrophys.* **686**, A236 (2024).

73 J Schou, and R Pedrys. *J. Geophys. Res. Planet.* **106**, 33309-33314 (2001).

74 RA Vidal, BD Teolis, and RA Baragiola. *Surf. Sci.* **588**, 1-5 (2005).

75 MP Collings, JW Dever, HJ Fraser, and MRS McCoustra. *Astrophys. Space Sci.* **285**, 633-659 (2003).

76 MP Collings, JW Dever, HJ Fraser, MRS McCoustra, and DA Williams. *Astrophys. J.* **583**, 1058-1062 (2003).